# Developing and evaluating a tutorial on the double-slit experiment

Ryan Sayer[1], Alexandru Maries[2], Chandralekha Singh[1]

[1]*Department of Physics and Astronomy, University of Pittsburgh, Pittsburgh, PA, 15260, USA*
[2]*Department of Physics, University of Cincinnati, Cincinnati, OH, 45221, USA*

**Abstract:** Learning quantum mechanics is challenging, even for upper-level undergraduate and graduate students. Interactive tutorials which build on students' prior knowledge can be useful tools to enhance student learning. We have been investigating student difficulties with the quantum mechanics pertaining to the double-slit experiment in various situations. Here we discuss the development and evaluation of a Quantum Interactive Learning Tutorial (QuILT) which makes use of an interactive simulation to improve student understanding. We summarize common difficulties and discuss the extent to which the QuILT is effective in addressing them in two types of courses.

## I. INTRODUCTION

According to a poll of Physics World readers, the interference of single electrons in Young's double slit experiment (DSE) is "the most beautiful experiment in physics" [1]. The beauty of this experiment comes from its powerful illustration of the quantum nature of very small particles. The experimental setup is useful for helping students learn about foundations of quantum mechanics, including the wave-particle duality of a single particle, the probabilistic nature of quantum measurements, wave function collapse upon measurement, etc. It illustrates how information about which slit a particle went through, or "which-path" information (WPI), can potentially destroy the interference pattern on the distant screen when a large number of single particles are sent [2,3]. Prior research shows that many students struggle with these unfamiliar quantum-mechanical concepts [4-8].

Here we summarize the development and evaluation of the DSE Quantum Interactive Learning Tutorial (QuILT) [7-8]. We analyze data collected from undergraduate and graduate students and determine the extent to which the QuILT was effective at addressing common student difficulties with the DSE.

## II. DEVELOPMENT OF THE QUILT

The development of the DSE QuILT started with an investigation of student difficulties. The QuILT strives to help students build on their prior knowledge and it addresses common difficulties. A preliminary version of the QuILT was developed based upon a cognitive task analysis of the underlying concepts and knowledge of common student difficulties and then iterated several times with five physics faculty members to ensure that they agreed with the content and wording. It was also administered to advanced undergraduate students in individual think-aloud interviews to ensure that the guided approach was effective, the questions were unambiguously interpreted, and to better understand the rationale for student responses. During these semi-structured interviews, students were asked to "think aloud" while answering the questions. Students first read the questions on their own and answered them without interruptions except that they were prompted to think aloud if they were quiet for a long time. After students had finished answering a particular question to the best of their ability, they were asked to further clarify and elaborate issues that they had not clearly addressed earlier. Modifications and improvements were made based upon the student and faculty feedback. A total of over 70 hours of individual interviews were conducted during the development and assessment phases of the DSE QuILT.

The DSE QuILT uses a guided approach to learning and accounts for common student difficulties (which will be summarized in Section IV). It consists of a pre-test, a warm-up, a main tutorial, an associated homework component, and a post-test. The pre-test is administered before students work on the warm-up, main tutorial and homework and then an identical post-test is administered after students work on all of the parts of the QuILT. The warm-up serves as a review of the de Broglie relation and the wave nature of particles as manifested in the DSE.

The main tutorial makes use of an interactive computer simulation in which students can manipulate the DSE setup and observe the resulting pattern. As they work through the tutorial in small groups in class, students are asked to predict the pattern that will appear on the screen in a given situation and then use the simulation to check their predictions. They are then given an opportunity to reconcile the difference between their predictions and observations before proceeding further. Students are also provided checkpoints to reflect upon what they have learned and to make explicit connections with their prior knowledge. After working on the main tutorial, students work on a homework component to connect the conceptual and mathematical aspects of the DSE. The post-test is given after all other QuILT components.

## III. PRELIMINARY EVALUATION

The QuILT was administered to undergraduates in an upper-level quantum mechanics course ($N = 46$) and to first year physics graduate students in a mandatory semester-long TA training course which met for 2 hours each week ($N = 45$). The students were given the pre-test during one class period, after which they worked on the warm up and the main tutorial in groups. Whatever they could not finish in class (including the homework component), they finished individually at home and submitted them to the instructor. The post-test was administered during the next class period.

The undergraduates received full credit for taking the pre-test, the completed tutorial counted as a small portion of the homework grade, and the post-test was graded for correctness as a quiz. In addition, the undergraduates were aware that topics discussed in the DSE tutorial may also appear in future exams since it was part of the course material. The graduate students worked on the tutorial and pre/post-test in the TA training course to help them recognize the value of using the tutorial approach to learning physics. While the graduate students completed all of the DSE QuILT materials similar to the undergraduates, their work was graded for completeness instead of correctness since their course performance was graded as satisfactory or unsatisfactory in the TA training course.

**Pre/Post-Tests: Question 1** presents a DSE set-up with single electrons and asks students to describe a situation in which the introduction of a lamp between the slits and the screen close to the slits would destroy the interference pattern (although the electrons still arrive at the screen). A correct response mentions that the wavelength of the photons emitted from the lamp should be smaller than the separation between the slits (in order to localize the electron sufficiently close to one of the two slits so that when an electron arrives at the screen we have WPI about the slit the electron went through). **Questions 2 and 3** present the DSE using sodium (Na) atoms and ask students to write the number density at a point $x$ on the screen and describe the pattern observed after many atoms reach the screen. The wavelength of the photons emitted from the lamp is *significantly smaller* (Q2) or *significantly larger* (Q3) than the slit separation, while the intensity is such that each Na atom scatters off a photon on both questions (but still arrives at the screen). The correct patterns on screen for Q2 and Q3 are no interference and full interference, respectively (which may be reasoned using WPI). **Questions 4 and 5** repeat Q2-3, but now the intensity of the lamp has been decreased so that only half of the Na atoms scatter off the photons. The correct patterns are partial interference (only Na atoms that do not scatter a photon show interference) and full interference (scattering does not localize Na atoms sufficiently to give WPI), respectively. The parameters for the photons that scatter off the Na atoms in the DSE situations in Q2-5 are summarized in Table I.

**TABLE I**. Summary of relevant properties of photons from the lamp that interact with Na atoms in the DSE for questions 2-5.

|  | Short Wavelength | Long Wavelength |
|---|---|---|
| **Full Intensity** | Question 2 | Question 3 |
| **Half Intensity** | Question 4 | Question 5 |

**Concept-based Rubric**: Student performance on the pre- and post- tests was evaluated using a "holistic" rubric which was designed to assess student understanding of relevant concepts across multiple questions. For example, one goal of the DSE QuILT is to help students learn that changing the wavelength of the photons emitted by the lamp can alter the interference pattern. Q2-3 (as well as Q4-5) were graded together using the rubric shown in Table II. Here are one student's actual responses to Q2-3:

> Q2: $\frac{N}{2}(|\psi_1|^2 + |\psi_2|^2)$. *No interference, even distribution of photons*
> Q3: $\frac{N}{2}(|\psi_1|^2 + |\psi_2|^2)$. *Still no interference pattern, since photons give path info for each electron*.

The rubric was applied to these responses as follows: The student received 0 points for part 1 (no mention of wavelength is made), 1 point for part 2 (the pattern is correct for Q2 but not Q3), 0 points for part 3 (the two number densities are not different), 1 point for part 4 (the number density is correct for Q2 but not Q3), and 2 points for part 5 (both number densities are at least consistent with the patterns described), for a total of 4 points. More than 20% of the data collected were independently rated by two different researchers and the inter-rater reliability was excellent (more than 90% agreement).

**TABLE II**. Summary of rubric to evaluate student responses to Q2-3 and Q4-5, with a total of eight points possible for each pair.

| 1. Mention that photon wavelength is an important consideration in determining pattern. | +1, 0 |
|---|---|
| 2. Correctly interpret the effect of wavelength on the interference pattern. | +2, +1, 0 |
| 3. Found different number densities. | +1, 0 |
| 4. Number densities are correct. | +2, +1, 0 |
| 5. Number densities consistent with patterns | +2, +1, 0 |

## IV. STUDENT PERFORMANCE

The average scores on the pre-/post-tests for the undergraduate and graduate students are shown in Table III with normalized gains [9]. Also as noted, Q2 and Q3 were graded together according to the rubric previously described, as were Q4 and Q5. The average scores for Q1, Q2-3, and Q4-5 are shown in Table IV. Tables III and IV show that, on average, graduate students performed better than undergraduates in the pre-test, while the undergraduates performed better on the post-test. The QuILT was administered to both groups over a short time frame without any additional in-class instructions. Since other aspects of implementation were similar in both courses, one possible reason for the post-test score discrepancy is that, as noted earlier, the undergraduates were given grade incentives to learn from the QuILT while the graduate students were given the QuILT in a TA training course with no final exam and a pass/fail grading scheme. The graduate students may be less motivated to learn from the QuILT if there is no grade incentive to improve.

**TABLE III**. Average pre-test and post-test scores, *p*-values, and normalized gains for undergraduate (U) and graduate (G) students.

|   | Pre Avg. | Post Avg. | *p*-value | Norm. Gain |
|---|---|---|---|---|
| U | 23% | 95% | <0.001 | 0.94 |
| G | 44% | 73% | <0.001 | 0.52 |

**TABLE IV**. Average pre-test and post-test percentages and *p*-values on Q 1, 2-3, and 4-5 for undergraduate (U) and graduate (G) students.

|   | Q 1 | | Q 2-3 | | Q 4-5 | |
|---|---|---|---|---|---|---|
|   | Pre | Post | Pre | Post | Pre | Post |
| U | 16 | 94 | 34 | 97 | 19 | 95 |
| G | 47 | 68 | 49 | 83 | 35 | 69 |
| p | <0.001 | <0.001 | 0.018 | 0.005 | 0.023 | <0.001 |

**Student Difficulties:** A number of difficulties specific to the effect of wavelength and intensity of photons from the lamp on the resulting pattern will now be discussed.

**Difficulty Recognizing the Effect of Lamp Wavelength on the Interference Pattern**: Question 1 asks students to describe a situation in which the introduction of a lamp would destroy the electron interference pattern on the screen. Many students struggled with this question on the pre-test and provided a variety of responses, but post-test performance was better, as shown in Table V. The responses in Table V are categorized as follows:

**(A) Mention $\lambda < d$:** A correct response mentioned that the wavelength of the lamp's photons should be shorter than the separation between the slits (WPI is known for the electrons in this case). These students demonstrated that they understood the role of photon wavelength in determining whether we have information about which slit the particle passed through to reach the screen. Credit was also given to students who described how scattering via a photon localizes the particles while altering their momenta.

**(B) Mention "Which-path" Information:** At least half credit was given to any students who mentioned that if WPI is known from the scattered photons, then the interference pattern vanishes even if they did not explicitly describe the connection between WPI and the wavelength of the lamp's photons. Any response that mentioned WPI is counted in this category, even if it was included in another category.

**TABLE V.** Categorization of student responses to Q1 as a percent of total responses. (A) is correct and (B) is partially correct.

| Q 1 | (A) | (B) | (C) | (D) | (E) | (F) |
|---|---|---|---|---|---|---|
| U Pre | 9% | 13% | 33% | 20% | 20% | 9% |
| U Post | 91% | 80% | 0% | 0% | 0% | 0% |
| G Pre | 14% | 32% | 36% | 5% | 14% | 5% |
| G Post | 64% | 22% | 9% | 4% | 16% | 2% |

This is why the rows do not add up to 100%.

**(C) Scattering:** The most common response on the pre-test described any type of physical scattering of the electrons due to collisions with the photons destroying the interference pattern without mentioning the constraints on photon wavelength. One student stated: *"If scattering occurs enough between the lamp photons & the particles, they will completely convolute the interference pattern so it will no longer be visible. The screen will simply appear completely lit up"* Another student stated: *"The interference pattern will be destroyed if the lamp has high enough intensity to scatter off the electrons."*

**(D) Photon-electron Interference:** Several students (mostly undergraduates) described situations in which the wavelengths and phases of the photon and electron were aligned in such a way that the two would destructively interfere. For example, one student wrote: *"For destructive interference to occur the phase (scattering angle) between the photon and the electron must be such that maxima of the photon's wavelength correspond to minima of the electron's wavelength and vice versa."* Interviews suggest that students with these types of responses often invoked the principle of superposition as though the photon and electron were identical particles and the crest of one particle will cancel the trough of the other particle wave.

**(E) Other Responses:** Many responses in this category were too simplistic and did not fall into other categories. For example, one student stated: *"There will be an interference pattern when the light bulb is off. When the light bulb is on, there will not be interference."*

**(F) Incomplete or No Response:** This category also includes those who simply wrote "I don't know".

On the pre-test 9% of undergraduates and 14% of graduate students were able to correctly identify the photon wavelength condition for whether an interference pattern will form on the screen. On the post-test 91% of undergraduates and 64% of graduate students received full credit for their responses. These results demonstrate that the QuILT was effective in addressing their initial difficulties. The discrepancy between undergraduate and graduate students may partly be due to the fact that the graduate students, who did not get a letter grade for their work in the TA training course, were less engaged with the QuILT.

Student responses to questions 2-3 were considered together, as were questions 3 and 5 (rubric not included for this pair). The responses for these pairs were divided into the following six categories:

(A) Patterns and number densities are both correct.
(B) Patterns are correct, but not the number densities.
(C) Patterns are different and incorrect.
(D) Patterns are the same and incorrect.
(E) Other responses.

Student responses to Q2 and Q3 were scored together to determine whether students understood what will happen in the experiment if the wavelength of the photons emitted by the lamp is altered. For Q2, the wavelength is significantly smaller than the distance between the two slits (which impacts the interference pattern), while for Q3, the wavelength is significantly larger. The breakdown of the student responses to this question pair is shown in Table VI.

TABLE VI. Categorization of undergraduate and graduate student responses to Q2 and Q3 as a percent of total responses.

| Q 2,3 | (A) | (B) | (C) | (D) | (E) | (F) |
|---|---|---|---|---|---|---|
| U Pre | **2%** | <u>30%</u> | 26% | 20% | 0% | 22% |
| U Post | **91%** | <u>9%</u> | 0% | 0% | 0% | 0% |
| G Pre | **25%** | <u>5%</u> | 30% | 20% | 5% | 16% |
| G Post | **71%** | <u>2%</u> | 9% | 13% | 4% | 0% |

**(A) Patterns & Number Densities Correct:** Table VI shows that graduate students were more likely than undergraduates to respond correctly to question pair 2-3 on the pre-test (25% vs. 2%, respectively). On the post-test, however, 91% of undergraduates answered correctly compared to only 71% of the graduate students.

**(B) Only Patterns Correct:** Table VI shows that about 30% of undergraduate students on the pre-test had a correct qualitative understanding of the role of photon wavelength in question pairs 2-3 but did not know how to correctly write the number densities.

**(C) Patterns Different, Incorrect:** Students in this category understood that changing the wavelength of the photons should change the pattern observed on the screen, but were not sure what that change should be.

**(D) Patterns the Same, Incorrect:** Table VI shows that in the pre-test, 20% of undergraduate and graduate students did not realize that changing the photon wavelength from significantly smaller to significantly larger than the distance between the slits will alter the pattern observed on the screen. Interestingly, 13% of graduate students on the post-test maintained that the two patterns should be the same. They either did not think that changing the photon wavelength should affect the interference pattern, or did not make an effort to distinguish between the two situations.

**(E) Other Responses:** Some graduate students drew pictures that may or may not have represented interference patterns in the researchers' view, and a few of them wrote "Yes" or "No" for their responses without any elaboration.

TABLE VII. Categorization of undergraduate and graduate student responses to Q3 and Q5 as a percent of total responses.

| Q 3,5 | (A) | (B) | (C) | (D) | (E) | (F) |
|---|---|---|---|---|---|---|
| U Pre | **4%** | <u>24%</u> | 35% | 9% | 0% | 28% |
| U Post | **80%** | <u>14%</u> | 7% | 0% | 0% | 0% |
| G Pre | **14%** | <u>5%</u> | 34% | 18% | 5% | 25% |
| G Post | **49%** | <u>2%</u> | 40% | 4% | 4% | 0% |

**(F) Incomplete or No Response:** About 22% of undergraduates and 16% of graduate students did not fully respond on the pre-test, or simply wrote "I don't know."

**Difficulty Recognizing the Effect of Lamp Intensity on the Interference Pattern**: Question pair 3 and 5 present a situation in which the intensity of the lamp is altered while the wavelength of the photons is significantly larger than the distance between the slits such that scattering between the photons and atoms will not affect the pattern on the screen. Student responses to these questions were compared and categorized, as shown in Table VII. (Correct responses are in bold and partially correct are underlined).

In Table VII, responses in categories (A) and (B) indicate that many students understood that lamp intensity does not matter in this situation. While about 94% of undergraduates recognized this fact on the post-test, only about 51% of the graduate students did so.

As seen in category (C) of Table VII, about 35% of undergraduates on the pre-test did not realize that photons with wavelengths longer than the distance between the slits cannot alter the interference pattern, regardless of their intensity. Interestingly, 40% of graduate students made this mistake on the post-test compared to 34% on the pre-test.

The common difficulty with question pair 3-5 illustrates a powerful phenomenological primitive, i.e., if you change something in the input, it should change *something* in the output [10]. However, in this case, changing the intensity of the lamp has no effect on the pattern. This primitive is specifically addressed in the QuILT but more graduate students used this primitive on the post-test than on the pre-test, which suggests that many of them did not engage with the QuILT in a meaningful way, unlike the undergraduates.

## V. SUMMARY

We developed and evaluated a DSE QuILT. Comparison of pre- and post-tests indicates that the DSE QuILT was effective in improving students' understanding of relevant concepts. The undergraduates improved more from the pre- to the post-test than graduate students with average normalized gains of 0.94 and 0.52, respectively [9]. This dichotomy may partly be due to differences in grade incentives as discussed. The use of the primitive discussed suggests that we should further explore these issues in advanced courses, especially in quantum mechanics [10].

## VI. ACKNOWLEDGEMENTS

We thank the National Science Foundation for award PHY-1202909, and Klaus Muthsam for the simulation.[1] R. Crease, Phys. World **15,** 19 (2002).
[2] R. Feynman et al., *Feynman Lectures* (1989).
[3] J. Wheeler, in Mathematical Foundations of Quantum Theory, edited by A.Marlow (Acad. Press, N.Y., 1979).
[4] C. Singh, Am. J. Phys. **69**, 885 (2001).
[5] D. Zollman et al., Am. J. Phys**. 70**, 252 (2002).
[6] M. Wittmann et al., Am. J. Phys. **70**, 218 (2002).
[7] Marshman and Singh, 2014 PERC Proc, 175 (2015).
[8] Singh and Marshman, 2014 PERC Proc, 239 (2015).
[9] R. Hake, Am. J. Phys., **66**, 64 (1998).
[10] A. diSessa, in *Constructivism in the Computer Age*, eds. Forman, Pufall (Lawrence Erlbaum, Hillsdale, (1988).